\documentclass[preprint,aps,eqsecnum]{revtex4}
\usepackage{latexsym}
\begin{document}
%\draft
%\preprint{\vbox{\baselineskip=12pt}
%\rightline{}
%\rightline{hep-th/0512224}
%\title{Brane stability with covariant perturbations in warped background}
\title{Geometry of deformations of branes in warped backgrounds}
%\title{Extremality and stability of branes in warped backgrounds}
\author{Supratik Pal \footnote{Electronic address: {\em supratik@cts.iitkgp.ernet.in}
; Permanent address: Relativity and Cosmology Research Centre, Department of Physics, 
Jadavpur University, Kolkata 700032, India}
${}^{}$ 
and Sayan Kar \footnote{Electronic address: {\em sayan@cts.iitkgp.ernet.in}}
${}^{}$}
\affiliation{Department of Physics and Centre for
Theoretical Studies\\ 
Indian Institute of Technology\\ 
Kharagpur 721 302, India}
\vspace{.5in}

\begin{abstract}

The `braneworld' (described by the usual worldvolume action) 
is a D dimensional timelike surface embedded in a 
N dimensional ($N>D$) warped, nonfactorisable spacetime. We first address
the conditions on the warp factor required to have an extremal flat brane
in a five dimensional background.
Subsequently, we deal with normal deformations of such extremal branes. 
The ensuing Jacobi equations are analysed to obtain the stability condition. 
It turns out that to have a stable brane, the warp factor should have a 
minimum at the location of the brane in the given background spacetime. 
To illustrate our results we explicitly check the extremality
and stability criteria for a few known co-dimension one braneworld models. 
Generalisations of the above formalism for the cases of (i) curved
branes (ii) asymmetrical warping and (iii) higher co-dimension braneworlds
are then presented alongwith some typical examples for each. Finally, we
summarize our results and provide perspectives for future work along these
lines. 

\end{abstract}

\maketitle

\section{Introduction}

Models with extra dimensions have been around for quite some time now.
Beginning with Kaluza--Klein {\cite{mkk}}, unified theories and 
followed in the later
parts of the twentieth century by string theory {\cite{gsw}}, 
a large variety of models with extra dimensions have been proposed. 
However, till now, we
are yet to find any experimental verification of the existence of
such dimensions. Among recent developments, we have had the so--called
braneworld models, which are, primarily, of two types--flat spacetime models
(the large extra dimension scenario) {\cite{add}}
and the warped models {\cite{rs}}. 
In the latter, we find that the four dimensional
submanifold (3-brane) has a dependence on the extra dimension through
the so--called warp factor, which, in turn makes the background
line element nonfactorisable.

All of the currently popular braneworld scenarios assume that we live
on a 3--brane in a background spacetime. This viewpoint
leads us to the fact that our world is an {\em embedded} hypersurface 
(the 3--brane) in a five dimensional background (called the `bulk').
One may also view the 3--brane as a domain wall or a defect in spacetime.
An extensive literature exists on such domain walls in supergravity
theories {\cite{cvetic}.

The action for such an embedded hypersurface would therefore be
the usual `worldvolume' action (dimensionally extending the standard Nambu--Goto
area action for strings) \cite{gsw}. It is thus relevant to ask
whether such hypersurfaces are extremal (vanishing of the trace of
the extrinsic curvature, $K=0$) and if so, whether
they are stable against small normal deformations. The aim of this
article is to address these issues in the context of the warped
braneworld models {\cite{rs}}.

We first write down the condition for extremality mentioned above
for arbitrary embedded surfaces in a warped background geometry.
Then, for different types of warping, we investigate whether this
extremality condition holds good. Subsequently, we move on to
the so--called Jacobi equations (obtained by constructing a
second variation of the worldvolume action) which describe the
perturbative normal deformations of the embedded surface. 
The solutions for these equations address the question of stability
through the nature of the modes--oscillatory modes indicate stability
whereas growing modes confirm the existence of an instability.
We check these consequences for the various warped braneworld models 
with one extra dimension that exist in the literature. Having dealt
with the flat branes (induced metric being scaled Minkowski) we
move on to discuss two types of curved branes--the spatially flat
cosmological braneworld and the spherically symmetric, static one.
The extremality and stability criteria for such curved branes are
obtained for some special cases. 
Finally, we  
analyse various possible extensions (eg. asymmetrically warped branes
and higher co-dimension branes) and summarize our results in brief.

%%%%%%%%%%%%%%%%%%%%%%%%%%%%%%%%%%%%%%%%%%%%%%%%%%%%%%%%%%%%%%%%%%%%%%%%%%%

\section{Extremality and Perturbations on worldsheet}

We consider a D-dimensional hypersurface (worldsheet) embedded in a 
N-dimensional background spacetime. The coordinates of the background are
$X^{\mu} =X^{\mu} (\xi^a)$ where $\xi^a$ are the coordinates on the worldsheet. 
Here $a \equiv 1...D$ are the worldsheet indices, $\mu \equiv 1...N$ the
background indices and $i \equiv 1...(N-D)$ are the normal indices. 

With the aid of the orthonormal spacetime basis $(E^{\mu}_a, n^{\mu}_i)$ consisting of D
tangents and $(N-D)$ normals satisfying the orthonormality conditions
\begin{eqnarray}
g_{\mu \nu} E^{\mu}_{a} E^{\nu}_{b} &=& \eta_{a b}\\
g_{\mu \nu} n^{\mu}_{i} n^{\nu}_{j} &=& \delta_{i j}\\
g_{\mu \nu} E^{\mu}_{a}n^{\nu}_{i} &=& 0
\end{eqnarray}
the extrinsic curvature tensor components of the worldsheet is defined as

\begin{equation}
K^i_{ab} = - g_{\mu \nu} E^\alpha_a (\nabla_\alpha E^\mu_b) n^{\nu i}
\label{eq:a1}
\end{equation}
where $g_{\mu \nu}$ is the background metric and 
$\nabla_\mu$ are the covariant derivatives with respect
to the background coordinates.
The usual Nambu-Goto area action for the worldsheet is given by

\begin{equation}
S = -\lambda \int d^{D}\xi \sqrt{-\gamma}
\label{eq:a2}
\end{equation}
where  $\gamma_{ab}= g_{\mu \nu} X^\mu_{,a} X^\nu_{,b}$ are the 
induced metric coefficients on the worldsheet. 
The equation of motion (EOM) for the worldsheet 
in the higher dimensional background is obtained by extremising 
$S$ subject to the deformation {\cite {hawk, 1vari}}

\begin{equation}
X^\mu(\xi^a) \rightarrow X^\mu(\xi^a)+\delta X^\mu(\xi^a) 
\label{eq:a3}
\end{equation}

For a minimal surface, the first variation of the action
gives the EOM comprising of $(N - D)$ equations 
involving the trace of the extrinsic curvature {\cite {1vari}} :

\begin{equation}
\gamma^{a b} K^i_{a b} = 0
\label{eq:a4}
\end{equation}

It  is worth mentioning here that $(N - D)$ is the number referred to as the
co-dimension of the embedded worldsheet.

Unlike the first variation equation, the procedure of finding out the second variation
equation is a bit tedious. Still, for the sake of completeness, here
we provide the major steps of calculation.
Since only the motion transverse to the worldsheet is relevant (tangential
perturbations can be gauged away via reparametrisation invariance) one
can write {\cite {rel, def}}
\begin{equation}
\delta X^\mu = \Phi^i n^{\mu i} 
\end{equation}
in terms of the $(N - D)$ scalars $\Phi^i$. The second variation of the
action is given by

\begin{equation}
\delta^2 S = \lambda \int d^D\xi ~\left[g_{\mu \nu} \delta^{\mu} {\cal D}_\delta {\cal D}_a
(\sqrt{-\gamma}\gamma^{ab} E^{\nu}_b) \right]
\label{eq:a5a}
\end{equation}
where $\delta =\delta X^{\mu}\partial_{\mu}$ is the spacetime
vector field and the derivatives ${\cal D}_a, {\cal D}_\delta$ are defined
as ${\cal D}_a = X^{\mu}_{,a} \nabla_{\mu}$ and ${\cal D}_\delta =
\delta X^{\mu} \nabla_{\mu}$ respectively.
We next project the term inside square-bracket of Eq (\ref{eq:a5a}) onto the
unit normal $n^{\nu i}$, that results in the following equation
\begin{equation}
g_{\mu \nu} n^{\mu i} {\cal D}_\delta {\cal D}_a
(\sqrt{-\gamma}\gamma^{ab} E^{\nu}_b) = 0
\label{eq:a5b}
\end{equation}
and utilise the spacetime Ricci identity to obtain
\begin{equation}
{\cal D}_\delta {\cal D}_a (\sqrt{-\gamma}\gamma^{ab} E^{\nu}_b) =
{\cal D}_a{\cal D}_\delta (\sqrt{-\gamma}\gamma^{ab} E^{\nu}_b)
+  R E^{\mu}_a \delta_{\mu} (\sqrt{-\gamma}\gamma^{ab} E^{\nu}_b)
\label{eq:a5c}
\end{equation}
Further, with the help of the projection tensor
$h^{\mu\nu}= g^{\mu\nu}- n^{\mu i} n^{\nu i}$, we write
\begin{equation}
\gamma^{ab} g_{\mu \nu}\delta^{\mu}   (R E^{\alpha}_a \delta_{\alpha}) E^{\nu}_b =
\left[ R_{\mu\nu} n^{\mu i} n^{\nu j}
- R_{\mu\alpha\nu\beta} n^{\mu i} n^{\alpha k} n^{\nu j}
n^{\beta k}\right]\Phi^{i}\Phi^{j}
\label{eq:a5d}
\end{equation}
where the Riemann and Ricci curvature tensor components and the 
Ricci scalar are those of the background metric. 
Subsequently, we substitute the results of Eq (\ref{eq:a5c}) 
and (\ref{eq:a5d}) into Eq (\ref{eq:a5b}),
decompose the first term into three parts and analyse each part
by introducing the surface torsion 
$T^{ij}_a = g_{\mu \nu}({\cal D}_a n^{\mu i}) n^{\nu j}= -T^{ji}_a$.
As a result, the second variation (written as an action now) turns out to be

\begin{equation}
S=\frac{1}{2}{\delta}^2 S=\frac{1}{2}\int d^D\xi \sqrt{-\gamma}\left[\Phi^{k}
\Box \Phi^{i} -\Phi^{k}(M^2)^{i}_{j}\Phi^{j}\right ]
\label{eq:a5e}
\end{equation}
where, $\Box$ is the worldsheet d'Alembertian and we have absorbed
all the torsion terms by re-defining the worldsheet derivative
as $\tilde \nabla_a^{ij} = \nabla_a \delta^{ij} - T^{ij}_a$.
Also, we have defined the effective mass matrix $(M^2)^{i j}$ to be 

\begin{equation}
(M^2)^{i j} =
R_{\mu \alpha \nu \beta} n^{\mu i} n^{\alpha k} n^{\nu j} n^{\beta k}
- R_{\mu\nu} n^{\mu i} n^{\nu j}
-K^{i}_{ab} K^{ab j}
\label{eq:a6}
\end{equation}

Finally, 
the variation of the action with respect to $\Phi^k$ 
leads to the following equation {\cite {2vari}}
\begin{equation}
\Box \Phi^i - (M^2)^i_j \Phi^j = 0
\label{eq:a5}
\end{equation}

What turns out from the above equation is that
for a Minkowski brane, a negative
eigenvalue of $(M^2)^{i j}$ will lead to an instability.

Furthermore, for a $(N - 1)$ dimensional hypersurface (co-dimension 1) 
with a single normal vector, both the torsion and the total
projected Riemann tensor onto the normal vanish, and Eq (\ref{eq:a5}) 
reduces to {\cite {2vari}}
\begin{equation}
\Box \Phi + (R_{\mu \nu} n^{\mu} n^{\nu} +K_{ab} K^{ab}) \Phi = 0
\label{eq:a7}
\end{equation}

Equations (\ref{eq:a4}) and (\ref{eq:a7}) are the two guiding principles in
discussing the extremality and stability of branes. In a nutshell, in
order that the worldsheet be extremal and stable, the extrinsic curvature tensor
has to be traceless and the effective mass matrix needs to have positive
eigenvalues.

%%%%%%%%%%%%%%%%%%%%%%%%%%%%%%%%%%%%%%%%%%%%%%%%%%%%%%%%%%%%%%%%%%%%%%%%%%%%%%%

\section{Schematics of warped spacetimes}

Let us now switch over to a brief discussion of the geometry of
warped extra dimensions.
We consider the by-now well-known Randall-Sundrum type {\cite {rs}} 
braneworlds where 
our 4D observable universe (called a brane) is a hypersurface embedded 
in a 5D geometry (with a bulk negative cosmological constant) where for the RS1 
two brane model (branes at 0 and $\pi r_c$ in the extra dimension) 
the extra dimension constitutes an 
$S^1/Z_2$ orbifold and for RS2 the extra dimension is infinite. The various 
parameters in RS1 (brane tensions, bulk cosmological constant, curvature scale)
are fine-tuned in such a way that the effective
brane cosmological constant $\Lambda_4$ turns out to zero. The tensions
on the so-called visible (at $\sigma = \pi r_c$) and hidden (at $\sigma=0$)
branes are of equal magnitude but of negative and positive signs respectively. 
Subsequently, a number of
thin and thick brane models {\cite {kanti, local, keha, eva, sg, phan, giov1,
giov2, tach, giov3}}
have been proposed by introducing different forms of bulk 
scalars and brane tensions, that can 
play the role of a bulk cosmological constant.
Proposals with $Z_2$-asymmetry also exist {\cite {z2as}}.

In this so-called warped geometric setup where the curvature is there
in the bulk but the brane is flat, the 5-dimensional background metric can be 
expressed as 
\begin{equation}
d S_5^2 = e^{2 f(\sigma)} \eta_{\mu \nu} d X^{\mu}d X^{\nu} + d\sigma^2
\label{eq:b1}
\end{equation}
where $f(\sigma)$ is the ``warp factor'' that incorporates the effect
of the extra-dimension $\sigma$ on the brane and renders the bulk
geometry nonfactorisable. 

The induced metric on the brane is given by 
\begin{equation}
d S_I^2 = e^{2 f(\sigma_0)}\eta_{a b} d \xi^{a} d \xi^{b}
\label{eq:b2}
\end{equation}
where $\sigma = \sigma_0$ is the location of the brane. The form of the
metric shows that at  $\sigma = \sigma_0$ the coordinates are re-scaled
so that we get back the usual 4D flat geometry. 

The basic questions we are now going to address are : what are the extremality
and stability
conditions in this warped geometry and which among the known braneworld solutions
available in the literature are extremal and stable?

%%%%%%%%%%%%%%%%%%%%%%%%%%%%%%%%%%%%%%%%%%%%%%%%%%%%%%%%%%%%%%%%%%%%%%%%%%%%%%%

\section{Extremality and stability conditions of branes}

We denote $X^{\mu} \equiv (t, x, y, z, \sigma)$ for background coordinates and
$\xi^{a} \equiv (\tau, x_1, y_1, z_1)$ for brane coordinates.
With the background metric given by Eq (\ref{eq:b1}) and the induced
metric of Eq (\ref{eq:b2}) obtained by the embedding at $\sigma = \sigma_0$,
the non-zero components of the normalised tangent vectors 
turn out to be

\begin{equation}
E^t_\tau = E^x_{x_1} = E^y_{y_1} = E^z_{z_1} = e^{- f(\sigma)}
\label{eq:c1}
\end{equation}
whereas the single normal vector is chosen as 
\begin{equation}
n^{\mu} = (0, 0, 0, 0, 1)
\label{eq:c2}
\end{equation}

With these expressions for tangent and normal vectors and keeping in 
mind that all the covariant derivatives are taken with respect 
to the background
coordinates,  the extrinsic curvature tensor turns out to be
diagonal with the components 

\begin{equation}
- K_{\tau \tau}= K_{x_1 x_1} = K_{y_1 y_1} = K_{z_1 z_1} = f^{\prime} (\sigma)
\label{eq:c3}
\end{equation}
where a prime denotes derivative with respect to $\sigma$. Further, for 
co-dimension 1, the components of the Ricci tensor need to be evaluated.
However, the only relevant component that has a non-trivial effect on the
perturbation equation is

\begin{equation}
R_{\sigma \sigma} = -4 \left [f^{''}(\sigma) + f^{' 2}(\sigma) \right] 
\label{eq:c4}
\end{equation}

The 1st variation equation (\ref{eq:a4}) requires a straightforward trace
calculation of the extrinsic curvature, which gives

\begin{equation}
f^{'}(\sigma_0) = 0
\label{eq:c5}
\end{equation}

All it tells is that for the brane to be extremal $f(\sigma)$ 
should have an extremum at the location of the brane.

Let us now proceed further to exploit the 2nd variation equation in 
order to find the nature of the extremum. With the expressions for
the Ricci tensor and extrinsic curvature given above, Eq (\ref{eq:a7}),
for an extremal brane, reduces to 

\begin{equation}
\left [\Box - 4 f^{''}(\sigma_0) \right] \Phi = 0
\label{eq:c6}
\end{equation}

Hence the stability requirement, {\em i.e.} a positive eigenvalue
of the effective mass matrix {\cite {2vari}}, in the present scenario,
reduces to
\begin{equation}
f^{''}(\sigma_0) > 0
\label{eq:c9a}
\end{equation}

Our demand can further be justified by the following argument.
Following standard textbook stuff 
we expect a wavelike solution for Eq (\ref{eq:c6})
of the form (in terms of the scaled coordinates 
$\xi_I^a = e^{f(\sigma_0)} \xi^a$)

\begin{equation}
\Phi = A_0 e^ {i (\omega \tau_I - \overrightarrow k . 
\overrightarrow \xi_I)}
\label{eq:c7}
\end{equation}
where $A_0$ is a constant. 
Since $\Phi$ is an arbitrary scalar, equations (\ref{eq:c6}) and
(\ref{eq:c7}) together give the oscillation frequency $\omega$ as 

\begin{equation}
\omega^2 = k^2 + 4 f^{''}(\sigma_0)
\label{eq:c8}
\end{equation}

Hence the sufficient condition for the frequency to be real,
and as a result, the brane to be stable is
\begin{equation}
f^{''}(\sigma_0) > 0
\label{eq:c9b}
\end{equation}
justifying our demand in Eq (\ref{eq:c9a}). 
The constraint on $f^{''}(\sigma_0)$ 
further reveals that the extremum is, in fact, a minimum.
However it should be mentioned that in no way this is a 
mandatory condition. The frequency can be real even when
$f^{''}(\sigma_0)$ is negative if the condition
\begin{equation}
k^2 > 4 |f^{''}(\sigma_0)| 
\label{eq:c10}
\end{equation}
is satisfied. Hence a positive value of  $f^{''}(\sigma_0)$
only guarantees brane stability.

Equations (\ref{eq:c5}) and (\ref{eq:c9a}) together provide
the stability and extremality conditions for branes in a warped background.
The bottomline is that to have a stable braneworld solution 
the warp factor should have a minimum and the brane has to be
located at the minimum.

Once the stability condition is fixed, one can now easily
find out the effect of perturbations by explicitly writing down the perturbed coordinates  
%in case the brane is found to be stable 

\begin{equation}
X^{' \mu} = X^{\mu} + \Phi n^{\mu} 
\end{equation}

In the present scenario, all but one coordinate remain
unperturbed. The lone perturbed coordinate is the extra dimension 

\begin{equation}
\sigma^{'} = \sigma_0 +  Re ~[A_0 e^{i (\omega \tau_I - 
\overrightarrow k . \overrightarrow \xi_I) }]
\end{equation}

Hence a stable flat brane in a warped background oscillates harmonically 
about the stable location $\sigma = \sigma_0$
with a frequency $\omega = \sqrt {k^2 + 4 f^{''}(\sigma_0)} $.

%%%%%%%%%%%%%%%%%%%%%%%%%%%%%%%%%%%%%%%%%%%%%%%%%%%%%%%%%%%%%%%%%%%%%%%%%%%%%%%%%%

\subsection{Checking the stability of some co-dimension one models}

The following table discusses the extremality and stability of a few
co-dimension one braneworld solutions available in the literature
{\cite {rs, kanti, local, keha, eva, sg, phan, giov1, giov2, tach, giov3}}.

%\subsection*{Table :}

\begin{center}
\begin{tabular}{|c|c|c|c|c|c|c|}

\hline \textbf{$f(\sigma)$}&\textbf{$f^{'}(\sigma_0)$}&\textbf{$f^{''}(\sigma_0)$}
&\text{inference} &\text {extremal} & \text {brane} & \textbf{$\omega^{2}$}\\
 & & & &\text {brane} & \text {location}  & \\ 
 & & & &\text {location} & \text {in models} & \\ 
 
\hline $- k |\sigma|$ \cite{rs}& $\neq 0$ & $< 0$ & unstable &-- & 0 &-- \\

\hline $\frac{1}{2} \ln [A e^{-b \sigma} + B e^{b \sigma}]$  \cite{kanti}& $= 0$ &$> 0$ 
& stable & $\frac{1}{2 b} \ln {A \over B}$ & $\sigma_1, - \sigma_2$ & $k^2 + 2 b^2$\\ 

%\hline $\ln (\sqrt{\Lambda} L \sinh {c -| \sigma| \over L})$ \cite{local} & 
%$\neq 0$ & $ < 0$ & unstable &-- & 0  &-- \\ 

\hline $ {c - |\sigma| \over L}  $ \cite{local} & $\neq 0$ & $< 0$ & unstable &-- & 0  &-- \\

%\hline $\ln (\sqrt{-\Lambda} L \cosh {c - |\sigma| \over L})$ \cite{local}& $= 0$ & $> 0$ 
%& stable & $\pm c$  & 0  & $k^2 + {4 \over L^2}$ \\ 

\hline $- \beta \ln \cosh^2(a \sigma)$ & $= 0$ & $< 0$ & stable if & 0  & 0 &
$k^2 - 12 a^2 \beta$ \\
$+ {\beta \over 2} \tanh^2(a \sigma) $\cite{keha}&&& ${k^2 > 12 a^2 \beta}$ & & & \\

\hline $ {1 \over 4} \ln ({4 \over 3}\sigma + c)+d $ \cite{eva}& $\neq 0$& $< 0$& unstable &-- 
& 0 &-- \\

\hline $  c \ln \mbox{sech}(b \sigma)$ \cite{sg, giov1, giov2}& $ = 0$& $< 0$&  stable if & 0  & 0 &
$k^2 - 4 b^2 c$ \\
&&&$k^2 > 4 b^2 c$ & & & \\

\hline $  c \ln \cosh (b \sigma)$ \cite{phan}& $= 0$ & $> 0$  & stable &0  & 0 & 
$k^2 + 4 b^2 c$ \\ 

\hline $ {a \over k} e^{- k |\sigma|} $ \cite{tach}& $= 0$& = 0 & extremal, &$\mp \infty$  
 & 0 &-- \\
&&& unstable & & & \\

\hline

\end{tabular}
\end{center}

Some comments on the above table are in order. From the geometrical analysis,
it is clear that very few braneworld solutions available in the literature
are extremal as well as stable. For example, a study on the models
in \cite{rs, local} reveal that any linear solution for the warp factor
is neither extremal nor stable, that re-establishes the 
instability of Randall-Sundrum
model. Also, the stability requirement imposes extra constraints on the
parameters involved in some of the braneworld models, {\em e.g,} \cite{keha, sg},
satisfying the inequalities. Further, for models in \cite{keha, sg, phan, giov1, giov2},
the brane locations in the respective models are their extremal and stable
locations whereas for the model in \cite{kanti}, the stability criteria
fix the brane location : $\sigma_1 = \frac{1}{2 b} \ln (A /B)$ or
$\sigma_2 = - \frac{1}{2 b} \ln (A /B)$.

To conclude, in order to have a physically acceptable braneworld solution,
one must check whether it is geometrically stable or not. Herein lies the 
importance of the present analysis.

%%%%%%%%%%%%%%%%%%%%%%%%%%%%%%%%%%%%%%%%%%%%%%%%%%%%%%%%%%%%%%%%%%%%%%%%%%%

\section{Generalisations of the formalism}

So-far we discussed the stability of the usual symmetrical warping 
for flat branes with co-dimension one (hypersurfaces). Apart from this,
several other braneworld models have been invoked to solve a few 
physical problems facing the earlier models and with the notion of extending
the bulk-brane scenario where curvature on the brane or more than
one extra (spacelike) dimensions are involved. 
However, as discussed earlier,
to have a valid braneworld solution one must check the stability of such
models as well. With this intention
we extend our formalism for the three different kinds of models, namely, 
(A) the braneworld models with curvature on the brane, (B) the asymmetrically
warped spacetimes with co-dimension 1 and (C) the warped models with co-dimension 2.

%\subsection{Branes with curvature}
\subsection{Curved branes in warped backgrounds}

Curved branes in warped backgrounds, reproducing the standard 4-dimensional
geometry at the location of the brane, is of much physical interest from
the cosmological and gravitational points of view.
In this subsection, we intend to find out the extremality and stability
conditions of these type of branes. More precisely, we explore 
two most important embedding geometries, one representing the spatially flat
Friedman-Robertson-Walker metric on the brane, another a spherically symmetric 
brane representing the Schwarzschild spacetime.

%%%%%%%%%%%%%%%%%%%%%%%%%%%%%%%%%%%%%%%%%%%%%%%%%%%%%%%%%%%%%%%%%%%%%

\subsubsection{Spatially flat FRW branes}

The background metric for which the 4-dimensional cosmology 
with a spatially flat FRW brane is recovered, can be written as \cite{ds}
(or in terms of transformed coordinates from \cite{cosmo})

\begin{equation}
dS_5 ^2 = e^{2 f(\sigma)} [-d t^2 +  a^2(t) (d x^2 + d y^2 + d z^2)] + d\sigma^2
\label{eq:d1}
\end{equation}
whereas the induced metric representing the brane is given by
\begin{equation}
dS_I ^2 = e^{2 f(\sigma_0)} [-d \tau^2 +  a^2(\tau) (d x_1^2 + d y_1^2 + d z_1^2)]
\label{eq:d2}
\end{equation}
where, as usual, $(t, x, y, z, \sigma)$ are the background coordinates and
$(\tau, x_1, y_1, z_1)$ are the brane coordinates.
Here $a(\tau)$ is the scale factor
for spatially flat cosmological models. Later on we shall choose it to be 
de-Sitter in order to study a specific, solvable case explicitly. 

The non-zero components of the normalised tangent vectors are
 
\begin{equation}
E^t_\tau =e^{- f(\sigma)} \hspace{.5 cm}; \hspace{.5 cm}
 E^x_{x_1} = E^y_{y_1} = E^z_{z_1} = e^{- f(\sigma)} a^{-1}(t)
\label{eq:d3}
\end{equation}
whereas the normal vector remains the same as for the flat brane. In this setup,
the components of the extrinsic curvature tensor $K_{a b}$ are 
$- K_{\tau \tau}= K_{x_1 x_1} = K_{y_1 y_1} = K_{z_1 z_1} = f^{\prime} (\sigma)$
and the relevant 
Ricci tensor component is $R_{\sigma \sigma}=-4 [f^{''}(\sigma) + f^{' 2}(\sigma)]$. 
It turns out that they are identical
 to their corresponding expressions for a flat
brane and although the other Ricci tensor components are different
from the corresponding expressions for a flat brane,
they do not affect the variation equations. As a result, the 1st variation 
equation involving the trace of $K_{a b}$ leads to the extremality
condition

\begin{equation}
f^{'}(\sigma_0) = 0
\label{eq:d4}
\end{equation}
which is the same as that of the flat brane.
And the 2nd variation equation gives  
\begin{equation}
\left [\Box - 4 f^{''} (\sigma_0) \right] \Phi = 0
\label{eq:d5}
\end{equation}

Notice
that the worldsheet d'Alembertian now includes curvature. Hence the stability 
requirement for a flat brane, {\em i.e,} a positive eigenvalue for the 
effective mass matrix, will not hold good in the present scenario.  
In order to find out the appropriate stability condition, we expand Eq (\ref{eq:d5})
for the above worldsheet metric (in terms of the scaled coordinates $ 
\xi_I^a = e^{f(\sigma_0)} \xi^a$), which reads

\begin{equation}
- \ddot \Phi + \frac{1}{a^2} \nabla^2 \Phi
-3 \frac{\dot a}{a} \dot \Phi - 4 f^{''}(\sigma_o) \Phi = 0
\label{eq:d6}
\end{equation}
where a dot denotes a derivative with respect to the scaled time
and $\overrightarrow \nabla$ is the gradient in terms of
the 3-space on the brane.
The solution of the above equation is of the form $\Phi = A(\tau_I) 
e^{-i \overrightarrow k . \overrightarrow \xi_I}$,
which, by transformation of the variable $A(\tau_I) = F(\tau_I) G(\tau_I)$, 
turns out to be

\begin{equation}
\Phi =  a^{- 3/2} F(\tau_I) ~e^{-i \overrightarrow k . \overrightarrow \xi_I}
\label{eq:d7}
\end{equation}
with the function $F(\tau_I)$ satisfying the differential equation

\begin{equation}
\ddot F + \left[-\frac{3}{2} \frac{\ddot a}{a} - \frac{3}{4} (\frac{\dot a}{a})^2 
+\frac{k^2}{a^2} + 4 f^{''}(\sigma_o)\right] F= 0
\label{eq:d8}
\end{equation}
 Hence all one has
to do is to solve for $F(\tau_I)$ with appropriate scale factor $a(\tau_I)$ 
and search for a decaying time-dependence of $\Phi$, 
that will establish brane stability.

Let us now explicitly perform the stability analysis 
for a de-Sitter brane for which $a(\tau_I) = e^{H \tau_I}$ ($H$ = Hubble constant).
Here Eq (\ref{eq:d8}) takes the form

\begin{equation}
\ddot F + \left[4 f^{''}(\sigma_o) - \frac{9}{4}H^2 + k^2 e^{-2H \tau_I} \right] F= 0
\label{eq:d9}
\end{equation}
which has a Bessel function solution 
 
\begin{equation}
F(\tau_I) = C_1 ~J_{-\nu} (\frac{k}{H} e^{-H \tau_I}) 
+ C_2 ~J_{\nu}(\frac{k}{H} e^{-H \tau_I}) 
\label{eq:d10}
\end{equation}
where the order of the Bessel function is defined by
$\nu = \frac{1}{H}\sqrt{\frac{9}{4} H^2 -4 f^{''}(\sigma_0)}$
and  $C_1$, $C_2$ are two arbitrary constants.
With this, the expression for $\Phi$ turns out to be

\begin{equation}
\Phi = e^{-\frac{3}{2} H \tau_I} \left[C_1 ~J_{-\nu} (\frac{k}{H} e^{-H \tau_I}) 
+ C_2 ~J_{\nu}(\frac{k}{H} e^{-H \tau_I}) \right]  
e^{-i \overrightarrow k . \overrightarrow \xi_I}
\label{eq:d11}
\end{equation}

The exponential term in the above expression decays with the 
worldsheet time whereas the Bessel functions are oscillatory. As a result,
$\Phi$ turns out to be a decaying function of the worldsheet time, as desired
by the stability criteria. 

An extreme case is if $\nu$ happens to be an integer ($= n$) {\em {i.e,}}  
when $4 f^{''}(\sigma_0) = (9/4 - n^2) H^2$. In that case, 
Eq (\ref{eq:d9}) has two linearly independent solutions, one a Bessel
function of integer order $n$, another a Neumann function,  

\begin{equation}
F(\tau_I) = C_1 ~J_{n} (\frac{k}{H} e^{-H \tau_I}) 
+ C_2 ~Y_{n}(\frac{k}{H} e^{-H \tau_I}) 
\end{equation}
The Neumann function being divergent, the solution involving it is ruled
out by the stability criteria.

We further notice that the order of the Bessel function has to be real. This gives the
following stability condition for a de-Sitter brane
\begin{equation}
f^{''}(\sigma_0) \leq \frac{9}{16} H^2 
\label{eq:d12}
\end{equation}

In brief, a de-Sitter brane is stable only if the warp factor at the
location of the brane has certain upper bound determined by the parameter $H$.

The most popular de-Sitter brane embedded in  warped background of the form
of Eq (\ref{eq:d1}) has the warp factor \cite{ds}

\begin{equation}
f(\sigma) = \ln \left(\sqrt{\Lambda} L \sinh {c -| \sigma| \over L}\right)
\label{eq:d13}
\end{equation}

Clearly, $f^{'}(\sigma) \neq 0$ for any value of $\sigma = \sigma_0$.
Hence the brane has neither a minimum nor a maximum anywhere. So this is not an
extremal brane. As a direct consequence, one cannot have any stable
location for the brane too.

Even though, we have explicitly worked out the case for a de--Sitter, it
should be clear that for other types of scale factors one can try to
understand the extremality and stability of the corresponding cosmological
brane using the formalism discussed above, though analytic solutions
of the Jacobi equations may not always be necessarily available.

%%%%%%%%%%%%%%%%%%%%%%%%%%%%%%%%%%%%%%%%%%%%%%%%%%%%%%%%%%%%%%%%%%%%%%%

\subsubsection{Spherically symmetric, static branes}

For a spherically symmetric, static brane embedded in a warped background, 
the background metric  can be expressed as \cite{hawkbh}

\begin{equation}
dS_5 ^2 = e^{2 f(\sigma)} [-e^{2 \nu(r)}d t^2 + e^{2 \lambda(r)} d r^2 
+ r^2 d \Omega^2] + d\sigma^2
\label{eq:dd1} 
\end{equation}
with the induced metric of the form
\begin{equation}
dS_I ^2 = e^{2 f(\sigma_0)} [-e^{2 \nu(r_1)}d \tau^2 + e^{2 \lambda(r_1)} d r_1^2 
+ r_1^2 d \Omega_1^2]
\label{eq:dd2} 
\end{equation}

Here the non-trivial normalised tangent vectors are

\begin{equation}
E^t_\tau =e^{- f(\sigma) - \nu(r)}, ~E^r_{r_1} = e^{- f(\sigma) -\lambda(r)},
~E^\theta_{\theta_1} =  \frac{e^{- f(\sigma)}}{r}, ~E^\phi_{\phi_1} = \frac{e^{- f(\sigma)}}
{r \sin\theta}
\end{equation}
which give rise to the extrinsic curvature tensor components as
$- K_{\tau \tau}= K_{r_1 r_1} = K_{\theta_1 \theta_1} = K_{\phi_1 \phi_1} = f^{\prime} (\sigma)$
and the Ricci tensor component $R_{\sigma \sigma}$ is $-4 [f^{''}(\sigma) + f^{' 2}(\sigma)]$. 
As before, the perturbation equations are independent of the other Ricci tensor components. 
Consequently, the 1st and 2nd variation equations resemble
Eq (\ref{eq:d4}) and Eq (\ref{eq:d5}) respectively. Here also
the first equation provides us with the standard extremality condition
but the second equation needs to be further studied in order to obtain the
stability criteria. Considering radial perturbation $\Phi = \Phi(\tau_I, ~r_I)$,
we explicitly write down the equation
in terms of the scaled worldsheet coordinates $\xi_I^a = e^{f(\sigma_0)} \xi^a$.

\begin{equation}
-\frac{\partial^2 \Phi}{\partial \tau_I^2} 
+ e^{2 (\nu - \lambda)} \frac {\partial^2 \Phi}{\partial r_I^2}
-  e^{2 (\nu - \lambda)} \left[\frac{2}{r_I}- \frac {d (\lambda -\nu)}{d r_I} \right] 
\frac {\partial \Phi}{\partial r_I} - 4 f^{''}(\sigma_0) e^{2 \nu} \Phi = 0 
\label{eq:dd3} 
\end{equation}

Its solution is of the form $\Phi = e^{i \omega \tau_I} B(r_I)$
which, by re-writing the variable $B(r_I)$ as $B(r_I) = F(r_I) G(r_I)$, becomes

\begin{equation}
\Phi =  \frac{e^{(\lambda - \nu) /2}}{r_I} ~F(r_I) ~e^{i \omega \tau_I} 
\label{eq:dd4} 
\end{equation}
where $F(r_I)$ now satisfies the equation

\begin{equation}
\ddot F +\left[\frac{1}{2} (\ddot\lambda - \ddot\nu)
-\frac{1}{4}(\dot\lambda - \dot\nu)^2 
-\frac{1}{r} (\dot\lambda - \dot\nu) 
+ \omega^2 e^{2(\lambda - \nu)} -4 f^{''}(\sigma_0) e^{2\lambda} \right] F = 0  
\label{eq:dd5} 
\end{equation}

Here a dot denotes a derivative with respect to $r_I$.
 Once again the stability
will be established if $\Phi$ is found to be a decaying function of 
the worldsheet radial distance $r_I$. 
%for specific expressions for $\nu(r_I)$ and $\lambda(r_I)$.

Specifically, the field outside a spherically symmetric gravitating body is given by
the  exterior Schwarzschild metric, for which the above equation
reduces to

\begin{equation}
\ddot F + \left[\frac{\frac{M^2}{r_I^4} + \omega^2}{ (1- \frac{2M}{r_I})^2}
- \frac{4 f^{''}(\sigma)}{(1- \frac{2M}{r_I})}  \right] F = 0  
\label{eq:dd6} 
\end{equation}

At a distance much larger than its Schwarzschild radius ($r_I \gg 2M$),
that is relevant for most of the gravitating bodies since their Schwarzschild radii
lie well within the actual radii, Eq (\ref{eq:dd6}) can be recast as

\begin{equation}
- \ddot F - \frac{4 M [\omega^2 - 2 f^{''}(\sigma_0)]}{r_I} F 
=  [\omega^2 -4 f^{''}(\sigma_0)] F 
\label{eq:dd7} 
\end{equation}
that looks like the well-known Hydrogen atom problem with 
$V(r_I) = - \frac{4 M}{r_I}[\omega^2 - 2 f^{''}(\sigma_0)]  < 0$.
To have a stable brane, one has to search for
the $l = 0$ bound state solutions, for which $E = \omega^2 -4 f^{''}(\sigma_0)
< 0$. Hence in the lowest order \cite{qm}

\begin{equation}
F(r_I) = R_{10}(r_I) = N e^{-k r_I} ~_1F_1 (0, ~2, ~2k r_I)
\label{eq:dd8} 
\end{equation}
where $~_1F_1 (a, ~b, ~z)$ is the confluent hypergeometric function.
Consequently, 

\begin{equation}
\Phi = N \frac{1}{r_I \sqrt{1 - \frac{2 M}{r_I}}} 
~e^{-k r_I} ~_1F_1 (0, ~2, ~2k r_I) ~e^{i \omega \tau_I}
\label{eq:dd9} 
\end{equation}
which, as desired, decays with the worldsheet radial distance.
We see that in order to satisfy both the conditions  $V(r_I) < 0$ and $E < 0$
simultaneously, the restriction on $f^{''}(\sigma_0)$ is

\begin{equation}
\frac{\omega^2}{4} < f^{''}(\sigma_0) < \frac{\omega^2}{2}
\label{eq:dd10} 
\end{equation}
that is the required stability condition for a Schwarzschild brane.
Since $\omega^2$ is positive definite, the Schwarzschild brane will be stable
if the warp factor has a minimum at the brane location, with the value
of $f^{''}(\sigma_0)$ lying within a fixed range given by the above equation.

A second linearly independent solution for Eq (\ref{eq:dd7}) can be obtained
 by taking note on the fact that the first solution behaves like
$F(r_I) = R_{10}(r_I) \sim e^{-2 k r_I}$ \cite{qm}. 
With the help of the Wronskian \cite{arfk}, the second solution 
is found to be

\begin{equation}
F_2(r_I) \sim \frac{e^{2 k r_I}}{4 k}
\end{equation}
which is clearly diverging and, as a result, is ruled out by the 
stability criteria. Hence, the only acceptable solution for $\Phi$ is 
provided by Eq (\ref{eq:dd9}).

As for the cosmological case, we have, in the above discussion, obtained the
Jacobi equations for generic, static, spherically symmetric branes, which
can always be used to understand extremality and stability in cases
other than Schwarzschild.

%%%%%%%%%%%%%%%%%%%%%%%%%%%%%%%%%%%%%%%%%%%%%%%%%%%%%%%%%%%%%%%%%%%%%%%%

\subsection{Asymmetrically warped spacetimes}

As mentioned before,
the asymmetrically warped spacetimes \cite{asym, asymsk} provide 
a generalisation of the usual warped extra dimensions. 
The generalisation is realted to having 
two different warp factors, one associated with time and another with 
the 3-space, both functions of the extra dimension alone.
Such a generalisation does have serious problems, \textit{e.g,} 
an apparent  Lorentz violation of the  4-dimensional equations \cite{asym}.
Despite these problems, it is interesting
to study different aspects of these models from the geometrical point of view
\cite{asymsk}.
Here we intend to study the issue of extremality and stability
for such asymmetrically warped spacetimes.

Considering the background metric to be of the form

\begin{equation}
d S_5^2 = - e^{2 f(\sigma)} d t^2 +  e^{2 g(\sigma)} (d x^2 + d y^2 + d z^2) + d\sigma^2
\label{eq:e1}
\end{equation}
and the induced metric on the brane as

\begin{equation}
d S_I^2 = - e^{2 f(\sigma_0)} d t^2 +  e^{2 g(\sigma_0)} (d x^2 + d y^2 + d z^2) 
\label{eq:e2}
\end{equation}
where  $f(\sigma)$ and   $g(\sigma)$ are the two warp factors, 
the non-zero components of the normalised tangent vectors read

\begin{equation}
E^t_\tau =e^{- f(\sigma)} \hspace{.5 cm}; \hspace{.5 cm}
 E^x_{x_1} = E^y_{y_1} = E^z_{z_1} = e^{- g(\sigma)}
\end{equation}
whereas the normal vector is chosen as the same as before. Subsequently the non-zero
components of the extrinsic curvature tensor turn out to be

\begin{equation}
K_{\tau \tau}= - f^{\prime} (\sigma) \hspace{.5 cm}; \hspace{.5 cm}
 K_{x_1 x_1} = K_{y_1 y_1} = K_{z_1 z_1} = g^{\prime} (\sigma)
\label{eq:e3}
\end{equation}
and the relevant Ricci tensor component is the $\sigma \sigma$ part

\begin{equation}
R_{\sigma \sigma} = -\left [(f^{''}(\sigma) + f^{' 2}(\sigma)) + 3 (g^{''}(\sigma) + g^{' 2}(\sigma)) 
\right] 
\label{eq:e4}
\end{equation}

With these the 1st and 2nd variation equations lead to the following 
conditions
for asymmetric warping :
\begin{eqnarray}
f^{'}(\sigma_0) + 3 g^{'}(\sigma_0) = 0 
\label{eq:e5}\\
f^{''}(\sigma_0) + 3 g^{''}(\sigma_0)  > 0
\label{eq:e6}
\end{eqnarray}
 
In this scenario we find no separate minima for the two different warp 
factors but the stability requirement imposes extra constraints on them.
As for example, for a constraint relation of the form $f(\sigma) = \nu g(\sigma)$
\cite{asymsk} equations (\ref{eq:e5}) and (\ref{eq:e6}) look a bit more familiar :

\begin{eqnarray}
(1 + 3/\nu) f^{'}(\sigma_0) = 0 
\label{eq:e5a}\\
(1 + 3/\nu) f^{''}(\sigma_0)  > 0
\label{eq:e6a}
\end{eqnarray}

The above equations reveal that the brane can be extremal if either of the two conditions 
$\nu = -3$, $f^{'}(\sigma_0) = 0$ is satisfied. But the first condition, namely, $\nu = -3$,
leads to a trivial solution where both $f(\sigma)$
and $g(\sigma)$ turn out to be constant. Hence for a stable extremal brane with linear
dependence between warp factors, the stability criteria demands that $f(\sigma)$
(and consequently, $g(\sigma)$)
should have a minimum on the brane location. With the solutions from \cite{asymsk} 

\begin{equation}
f(\sigma) = \frac{\nu}{\eta} \ln(\eta \sigma + C) \hspace{.5 cm}; \hspace{.5 cm}
g(\sigma) = \frac{1}{\eta} \ln(\eta \sigma + C)
\end{equation}

It can be verified that $f(\sigma)$  or $g(\sigma)$ has no minimum anywhere.
Hence, this type of asymmetrically warped vacuum
solutions are not stable.

%%%%%%%%%%%%%%%%%%%%%%%%%%%%%%%%%%%%%%%%%%%%%%%%%%%%%%%%%%%%%%%%%%%%%%%%%%%%%

\subsection{Branes with co-dimension 2}

Brane models with co-dimension 2 have several advantages : all the standard
model fields and gravity can be localised on the same brane \cite{6dloc} and 
the negative tension
brane (in the original RS models) is no longer required to solve the so-called `Hierarchy
problem' \cite{6dten}. In this setup we have two extra dimensions and 
two $\sigma$-dependent
warp factors, one associated with
the flat 4D part and another with the sixth dimension $\theta$. 
The background metric, 
now 6-dimensional, is taken as \cite{6dloc,6dten,6d} 

\begin{equation}
d S_6^2 = e^{2 f(\sigma)} (- d t^2 + d x^2 + d y^2 + d z^2) + d\sigma^2 + e^{2 g(\sigma)}
d \theta ^2
\label{eq:e7}
\end{equation}
and the induced metric representing the 4-dimensional flat brane as
\begin{equation}
d S_I^2 = e^{2 f(\sigma_0)} (- d t^2 + d x^2 + d y^2 + d z^2) 
\label{eq:e8}
\end{equation}

Obviously, the tangent vectors are the same as those in Eq  (\ref{eq:c1}) and, without loss of
generality, we make our choice of normal vectors, satisfying the normalisation conditions, as

\begin{equation}
n_1 = (0,0,0,0,1,0) \hspace{.5 cm}; \hspace{.5 cm} n_2 = (0,0,0,0,0, e^{-g(\sigma)})
\end{equation}

With this choice, $K_{a b}^{2}$ becomes identically zero and the 1st variation equation 
(\ref{eq:a4}) for $K_{a b}^{1}$ reveals that

\begin{equation}
f^{'} (\sigma_0) = 0
\label{eq:e9}
\end{equation}
that is to say that an extremal brane has to be located $\sigma = \sigma_0$,
 the extremum of $f(\sigma)$. 
This condition is identical to the one derived for co-dimension 1 scenario. 
It can be checked that for any other choice of the normal vectors, every 
component of $K_{a b}^{2}$ is related to the corresponding component of
$K_{a b}^{1}$ by $K_{a b}^{2} = F(\sigma) \times K_{a b}^{1}$,
%bears a linear dependance on $K_{a b}^{1}$, 
the trace of which gives no extra equation as such. Hence our
choice of normal vectors are indeed justified and sufficient for the present discussion.
One might have wondered why no constraint is there on the warp factor
associated with the 6th dimension $\theta$, even if the metric looks asymmetrically warped
in a sense that there are two different warp factors. 
The fact that the 4-dimensional part that represents the brane is associated with
a single warp factor independent of $\theta$
is the obvious answer to it. 

In order to calculate the effective mass matrix $(M ^ 2) ^{ij}$ for two extra dimensions, 
one requires both Riemann and Ricci tensors (along with the extrinsic 
curvature). 
The relevant components of these quantities are listed below.
 
\begin{eqnarray}
R_{\sigma \theta \sigma \theta} = R_{\theta \sigma \theta \sigma} = 
- e^{2 g(\sigma)} [g^{''}(\sigma) + g^{' 2}(\sigma)] \\
R_{\sigma \sigma} = - 4 [f^{''}(\sigma) + f^{' 2}(\sigma)] - [g^{''}(\sigma) + 
g^{' 2}(\sigma)]\\
R_{\theta\theta} = - e^{2 g(\sigma)}[g^{''}(\sigma) + g^{' 2}(\sigma) 
+ 4 f^{' }(\sigma) g^{' }(\sigma)]
\end{eqnarray}

With these Eq (\ref{eq:a5}) now reduces to two equations for two
scalars $\Phi ^{1}$ and $\Phi ^{2}$ :

\begin{eqnarray}
\left [\Box - 4 f^{''} (\sigma_0) \right] \Phi^{1} = 0
\label{eq:e10}\\ 
\left [\Box - 4 f^{'} (\sigma_0) g^{'} (\sigma_0) \right] \Phi^{2} = 0
\label{eq:e11}
\end{eqnarray}

The first equation immediately provides us with the stability condition  as before : 
the extremum of $f (\sigma)$ has to be a minimum at the location of the brane. 
On the other hand, $f^{'} (\sigma_0)$ being zero,
the second equation reveals that the perturbations about
$\theta$ are always stable. Once again the obvious reason behind it is that the  warp
factors are functionally independent of $\theta$.  

To study an example, let us analyze the model derived in the first reference of \cite{6d} where

\begin{equation}
f(\sigma) = \cosh^{4/5}(k \sigma) \hspace{.5 cm}; \hspace{.5 cm}
g(\sigma) = g_0 \frac{\sinh^{2}(k \sigma)}{\cosh^{6/5}(k \sigma)}
\label{eq:e12}
\end{equation}

It is found that $f(\sigma)$ has a minimum at $\sigma = 0$. As a
consequence, this braneworld model
with co-dimension 2 turns out to be both extremal and geometrically stable.

%%%%%%%%%%%%%%%%%%%%%%%%%%%%%%%%%%%%%%%%%%%%%%%%%%%%%%%%%%%%%%%%%%%%%%%%%%%

\section{Summary and outlook}

In this article, we have obtained the criteria under which an
embedded brane (in a warped background) can be extremal and stable against 
small normal perturbations. We have applied these criteria to
several different flat braneworld models in warped five dimensional 
backgrounds and found that quite a few of the well--known models do not 
seem to meet them. 

Furthermore, we have generalised our formalism to other related contexts 
such as symmetrically warped branes with curvature (cosmological branes and
static, spherically symmetric branes), asymmetrically warped braneworlds 
and models with codimension 2. For each of the above cases we have 
obtained the extremality and stability criteria and illustrated them 
with some examples. 

It must be mentioned that there can be different ways to
embed a lower dimensional manifold in a given higher dimensional background.
We have chosen the simplest (and widely used) embedding throughout--
more complicated embeddings giving rise to nontrivial and different 
induced metrics can yield quantitatively different results. 
Additionally, it is also important to state that it is not
always necessary for the action to be Nambu--Goto (which is, again the
simplest and widely used choice). There can be additional terms such as
rigidity corrections, terms involving extrinsic curvature etc. which
might change the extremality and stability criteria. 
Finally, an issue which has not been dealt here is that of
large deformations (eg. formation of cusps and kinks on the worldsheet)
which requires the use of the so--called generalised Raychaudhuri
equations{\cite{grc}}. We hope to address these issues in future investigations.

\end{document}